\documentclass[twocolumn,amsmath,amssymb,nobibnotes,nofootinbib,preprintnumbers]{revtex4-1}

\usepackage{graphicx}
\usepackage{bm}
\usepackage{xcolor}

\usepackage{amsmath}	
\begin{document}

\preprint{YITP-14-57}

\title{Scalar suppression on large scales in open inflation}

\author{Jonathan White}%
 \email{jwhite@yukawa.kyoto-u.ac.jp}

\author{Ying-li Zhang}%
 \email{yingli@yukawa.kyoto-u.ac.jp}

\author{Misao Sasaki}
 \email{misao@yukawa.kyoto-u.ac.jp}

\affiliation{Yukawa Institute for Theoretical Physics, Kyoto University,
Kyoto 606-8502, Japan}%

\date{\today}

\begin{abstract}
We consider two toy models of open inflation and investigate their
 ability to give a suppression of scalar power on large
 scales whilst also satisfying observational constraints on the spatial curvature
 of the universe.  Qualitatively we find that both models are indeed
 capable of fulfilling these two requirements, but we also see that effects
 of the quantum tunnelling must be carefully taken into account if we
 wish to make quantitative predictions.    
\end{abstract}

\pacs{98.80.Cq}
\maketitle

\section{Introduction}

Precision measurements of the cosmic microwave background (CMB) by the WMAP
and {\it Planck} collaborations are in very good agreement with what has
become the standard model of cosmology, namely inflation plus
$\Lambda$CDM \cite{Hinshaw:2012aka, Ade:2013zuv, Ade:2013uln}. There
are, however, some anomalies, such as a 5--10\% deficit in the
temperature power spectrum on large scales ($l\lesssim 40$)
\cite{Ade:2013kta}.  Whilst the statistical significance of this anomaly
is currently only $2.5$--$3\sigma$, the tension will be exacerbated if
the recent findings of the BICEP2 team -- who have reported a
tensor-to-scalar ratio of $r \sim \mathcal O(0.1)$ \cite{Ade:2014xna} --
are confirmed \cite{Bousso:2014jca,Smith:2014kka}.  The reason for this increased
tension is easy to understand: the temperature power spectrum on large
scales ($l\lesssim 100$) receives contributions from both scalar and
tensor perturbations, meaning that for a given observed power the scalar
contribution must be suppressed if there exists a non-negligible tensor
contribution.

Perhaps the simplest way in which a suppression on large scales can be
accommodated is to allow for a negative running of the spectral index,
defined as $\alpha_s\equiv dn_s/d\ln k$.  What one finds, however, is
that the required value of $\alpha_s$ is $\sim \mathcal O(0.01)$,
\cite{Ade:2014xna, Abazajian:2014tqa}, which is one or two orders of
magnitude larger than the running predicted in standard slow-roll models
of inflation.  Moreover, a large running also tends to spoil the fit to
data on small scales.  As such, it appears that a non-power-law scalar
spectrum may be preferred \cite{Abazajian:2014tqa,Hazra:2014aea}, namely
one that is able to give a localised suppression on large scales.  Note
that one can also try to alleviate tensions between the BICEP2 results
and those of {\it Planck} by allowing for a modified tensor spectrum
\cite{Smith:2014kka,Gong:2014qga,Gerbino:2014eqa}, but here we focus on
modifying the scalar spectrum.

In canonical single-field models of inflation the power spectrum takes
the simple form
\begin{equation}
 \mathcal{P}_{\mathcal R}(k) = \left.\frac{H^4}{4\pi^2\dot\phi^2}\right|_{k=aH},
\end{equation}
where the subscript $k=aH$ denotes the fact that the quantity should be
evaluated at the time when the scale $k$ left the Hubble horizon.  Given
the inverse dependence on $\dot\phi$, a natural way in which we might
seek to suppress the spectrum would be to require that the field be
rolling quickly during the period that large scales left the horizon,
i.e. during the first few observable e-foldings of
inflation.\footnote{Note that there are other possible mechanisms, see
e.g. \cite{Contaldi:2003zv,Cline:2003ve,Powell:2006yg,Kawasaki:2014fwa}.}
This mechanism for suppression has been considered in the literature,
see
e.g. \cite{Contaldi:2003zv,Jain:2008dw,*Jain:2009pm,Pedro:2013pba,Hazra:2014jka},
and in canonical single-field inflation models an enhancement of
$\dot\phi$ is associated with a steepening of the potential towards the
onset of inflation.  Whilst such a tuned steepening might seem
unnatural, this is not necessarily the case in the context of open
inflation \cite{Bousso:2013uia, Bousso:2014jca, Freivogel:2005vv,
Freivogel:2014hca, Linde:1998iw, Linde:1999wv}.

\begin{figure}
\begin{center}
\includegraphics[width = 0.9\columnwidth]{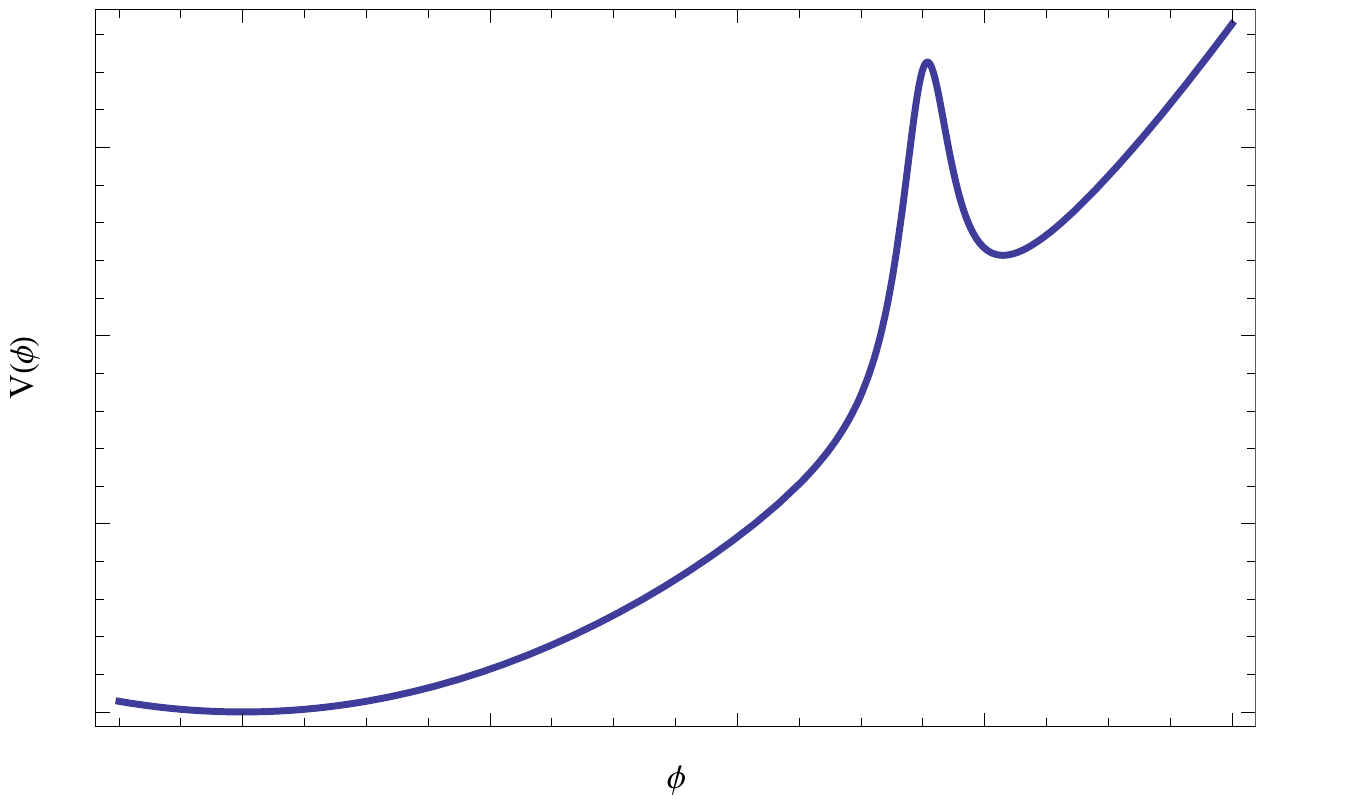}
\caption{\label{openpot}The general form of the potential associated with
 open inflation models.  Inflation is preceded by tunnelling from the
 false vacuum and the steepening of the potential in the vicinity of the
 barrier leads to a suppression of the scalar power spectrum.}
\end{center}
\end{figure}

Models of open
inflation involve potentials of the form shown in figure \ref{openpot},
and the last stage of observable inflation is preceded by tunnelling from
a false vacuum described by a Coleman-De Luccia (CDL) instanton
\cite{Coleman:1980aw}.  The universe nucleated in the tunnelling process
has negative curvature, i.e. it is open.  In general, CDL instantons require
$|V_{\phi\phi}|> H^2$ during the tunnelling process \cite{Jensen:1983ac}, which
is in contradiction with the requirement $|V_{\phi\phi}|\ll
H^2$ for slow-roll inflation.  As such, it is expected that between the end of
the tunnelling process and the onset of standard slow-roll inflation one has an
intervening stage of ``fast-roll'' inflation, during which the potential
becomes progressively less steep.  In order that observable signatures of
this fast-roll phase remain, we require that the ensuing stage of
slow-roll inflation does not last too long.  If the number of
e-foldings is too small, however, then the model becomes tightly
constrained by observational constraints on $\Omega_{{\rm K}}$ today.  The
compatibility of these two competing effects was discussed by Freivogel
{\it et al.} and Bousso {\it et al.}
\cite{Freivogel:2005vv,Freivogel:2014hca,Bousso:2013uia,Bousso:2014jca}.
It was found that models providing sufficient suppression of the
scalar spectrum whilst satisfying bounds on $\Omega_{{\rm K}}$ are possible if the
steepening of the potential towards the barrier is gradual enough.  

In this paper we revisit the toy models of open inflation
presented in \cite{Linde:1998iw} and \cite{Linde:1999wv}, and
investigate their ability to produce a suppression of the large-scale
scalar spectrum whilst evading observational constraints on $\Omega_{{\rm K}}$.     
In section \ref{open} we begin by giving a brief summary of open
inflation and present the relevant background and perturbation equations.  In
section \ref{models} we then analyse two of the toy models discussed in
\cite{Linde:1998iw}.  Based on the two toy models, in section
\ref{general} we then summarise more generally the nature of
suppression from open inflation models and how they can be constrained
using observations.  Finally we conclude in section \ref{conclusions}

\section{\label{open}Open inflation}

To describe the tunnelling process preceding single-field open inflation, we consider a class of model whose effective potential $V(\phi)$
has a local minimal at $\phi_0$ and global minimum at $\phi=0$ with
$V(0)=0$, as depicted in figure \ref{openpot}. The action is given by
\begin{eqnarray}\label{act}
\mathcal{S} =\int \sqrt{-g}
d^4x\left[\frac{R}{2}-\frac{1}{2}g^{\mu\nu}\partial_\mu\phi\partial_\nu\phi-V(\phi)\right]\,.
\end{eqnarray}
Note that here and throughout the paper we set $8\pi G=1/M_{{\rm Pl}}^2 = 1$.

Conventionally, the instanton solution is explored under the
assumption of $\mathcal{O}(4)$ symmetry, since it has been proved
that the $\mathcal{O}(4)$-symmetric solution gives the lowest value
of the Euclidean action for a wide class of scalar-field
theories~\cite{Coleman:1977th}. Hence, this solution is
exponentially favoured when calculating the corresponding tunnelling
probability. It is also reasonable to use this assumption with
gravity included~\cite{Coleman:1980aw}.
To be specific, the metric of an
$\mathcal{O}(4)$-invariant Euclidean spacetime can be written as
\begin{eqnarray}\label{Emetric}
ds_{\mathrm{E}}^2=d\xi^2+a_{\mathrm{E}}^2(\xi)\left(d\chi_{\mathrm{E}}^2+\sin^2\chi_{\mathrm{E}}d\Omega_2^2\right)\,.
\end{eqnarray}
The corresponding Euclidean background equations are given as
\begin{eqnarray}
&~&H_{\mathrm{E}}^2=
   \frac{1}{3}\left(\frac{\dot\phi^2}{2}-V  \right)+\frac{1}{a_{\mathrm{E}}^2}\,,\label{EFried}\\
&~&\ddot\phi+3H_{\mathrm{E}}\dot\phi-V_{\phi}=0\,,\label{Escalar}
\end{eqnarray}
where a dot denotes differentiation with respect to $\xi$ and
$H_{\mathrm{E}}\equiv\dot a_{\mathrm{E}}/a_{\mathrm{E}}$.  For the CDL instanton solution, we choose
$\xi=0$ to coincide with the centre of the nucleated $\mathcal O(4)$-symmetric
bubble, at which $\phi = \phi_{{\rm exit}}$, $a_{{\rm E}}= 0$ and $\dot
a_{{\rm E}}= 1$.  We further set $\xi = \xi_{{\rm F}}$ ($>0$) in the
false vacuum, where $\phi=\phi_{{\rm F}}$, $a_{{\rm E}}=0$ and $\dot
a_{{\rm E}} = -1$.

On making an analytic continuation to the Lorentzian space corresponding
to our open universe, the metric takes the form \cite{Sasaki:1994yt,Sasaki:1994yt,Yamauchi:2011qq,Sugimura:2013cra}
\begin{align}\label{Rmetric}
ds^2&=-dt^2+a^2(t)\left(d\chi^2+\sinh^2\chi d\Omega^2\right)\nonumber\\
&=a^2(\eta)\left(-d\eta^2+d\chi^2+\sinh^2\chi d\Omega^2\right)\,,
\end{align}
where $t = -i\xi$, $a = -ia_{{\rm E}}$, $\chi = i\chi_{{\rm E}}$ and $a
d\eta = dt$, and
the background equations of motion become
\begin{gather}
H^2 = \frac{1}{3}(\frac{1}{2}\dot\phi^2+V) + \frac{1}{a^2},\label{bgFe} \\
\ddot\phi + 3H\dot\phi + V_\phi = 0,\label{bgphiem}
\end{gather} 
where a dot now denotes taking the derivative with respect to $t$.  The
initial conditions coming from the instanton solution are $\phi =
\phi_{{\rm exit}}$, $\dot\phi = 0$, $a = 0$ and $\dot a = 1$.  Using these
equations of motion and initial conditions, we then need to solve for the
ensuing inflationary dynamics.

Locating the
observer at the centre of the spherical coordinates, the only tensor
modes that contribute to the CMB temperature power spectrum are the even-parity modes, which can be expressed as
\begin{equation}
 \delta g_{ij} = a^2t_{ij};\quad t_{ij} = \sum\hat b_{plm} U_p(\eta)Y_{ij}^{(+)plm}+\rm{h.c.}.
\end{equation}
Here $\hat b_{plm}$ are the annihilation operators, $Y_{ij}^{(+)plm}$
are the even-parity tensor harmonics on a unit 3-hyperboloid and
$\eta$ is the conformal time. The
$U_p(\eta)$ are found to satisfy \cite{Garriga:1998he} 
\begin{equation}\label{Ueom}
 U_p^{\prime\prime} + 2\mathcal H U_p^\prime + (p^2+1)U_p = 0,
\end{equation}
where a prime denotes $d/d\eta$ and $\mathcal H = a^\prime/a$.  

The important scalar quantity is the comoving curvature perturbation
$\mathcal R_c^p$, but this is more conveniently expressed in terms of a
new variable $\bm q^p$ as
\begin{equation}\label{Rqrel}
 \mathcal R_c^p = -\frac{\dot\phi}{2} \bm q^p -
		    \frac{H}{a\dot\phi^2}\frac{d}{dt}\left(a\dot\phi\bm q^p\right).
\end{equation}
The variable $\bm q^p$ then satisfies the simple equation
\begin{equation}\label{qeom}
 \bm q^{p\prime\prime} - \left[\frac{\phi^{\prime 2}}{2} + \phi^\prime\left(\frac{1}{\phi^\prime}\right)^{\prime\prime}-(p^2+4)\right]\bm q^p = 0.
\end{equation}

The scalar and tensor power spectra are given in terms of $\mathcal
R_c^p$ and $U_p$ as 
\begin{equation}
 \mathcal P_{\mathcal R} = \frac{p^3}{2\pi^2}\left|\mathcal R_c^p\right|^2\quad\mbox{and}\quad\mathcal P_{T} = \frac{p^3}{2\pi^2}\left|U_p\right|^2
\end{equation}
respectively.  In this work
we fit the power spectra using the following functional forms:
\begin{align}\label{scalar_mag}
 \mathcal P_{\mathcal R} &=
  \left(\frac{H^2}{2\pi\dot\phi}\right)^2_{t=t_{\mathcal R,p}}\frac{\cosh(\pi
  p)+\cos(\delta_p)}{\sinh(\pi p)}\frac{p^2}{c_1^2+p^2},\\
\mathcal P_{T} &=4\left(\frac{H}{2\pi}\right)^2_{t=t_{T,p}}\frac{\cosh(\pi
 p)-1}{\sinh(\pi p)}\frac{p^2}{c_2^2+p^2},\label{tensor_mag}
\end{align}  
where the subscripts $t=t_{\mathcal R,p}$ and $t=t_{T,p}$ indicate that
the quantity should be evaluated at the horizon-crossing time for scalar
and tensor modes, respectively.  The phase $\delta_p$ is irrelevant for
large $p$ and behaves as $\delta_p - \pi \propto p$ for $p\rightarrow
0$.  In fact, we take $\delta_p = \pi$ throughout, which corresponds to
the case where maximal suppression is achieved.  The above expressions
for $\mathcal P_{\mathcal R}$ and $\mathcal P_{T}$ are based on analytic
approximations derived in \cite{Yamamoto:1996qq,Garriga:1998he}.  We can
see that they differ from the standard expressions by the $
p$-dependent ``suppression factors,'' which tend to unity for large $
p$ and $\pi p^3/(2c_{1,2}^2)$ for $p \rightarrow 0$.  This $
p$-dependent suppression is distinct from that associated with the
fast-roll dynamics, and reflects the systems memory of the quantum
tunnelling that preceded inflation \cite{Linde:1999wv}.  The parameters $c_1$ and $c_2$
determine the scales at which this additional suppression becomes active
for the scalar and tensor modes, respectively.  Under the weak
back-reaction approximation, the spectra were found to take the above
form with $c_1 = 1$ and $c_2 =1$ \cite{Yamamoto:1996qq,Garriga:1998he}.
In Appendix C of \cite{Garriga:1998he}, similar expressions were also
found for an analytically soluble model, but with $c_1 = 2$ and $c_2 =
1$.  In this paper we take the two parameters $c_1$ and $c_2$ as free
parameters that we use to fit the numerical results of
\cite{Linde:1999wv}.

In order to determine the horizon-crossing conditions we return to the
equations of motion.  Firstly, for $U_p$, we see from \eqref{Ueom} that
horizon crossing takes place when $a^2H^2 = p^2 + 1$, as in the standard
case.  In order to determine $t_{\mathcal R,p}$, we first need to use
\eqref{Rqrel} and \eqref{qeom} to find an equation of motion for
$\mathcal R_c^p$.  We find\footnote{This equation can also be derived using the action for $\mathcal R_c$ given in Appendix B of \cite{Garriga:1997wz}.} 
\begin{equation}
 \mathcal R_{c}^{p\prime\prime} + 2A(\eta,p)\mathcal R_c^{p\prime} +
  B(\eta,p)\mathcal R_c^p = 0, 
\end{equation} 
with 
\begin{align}
 A(\eta,p) &= \frac{k^2\mathcal H^2\frac{\mathcal
 H}{a\phi^\prime}\left(\frac{a\phi^\prime}{\mathcal H}\right)^\prime -
 \frac{\phi^{\prime 2}}{2}\mathcal H}{k^2\mathcal
 H^2-\frac{\phi^{\prime 2}}{2}},\\
B(\eta,p) &= \frac{k^4\mathcal H^2 + \left(1+\frac{\mathcal H^2}{\phi^{\prime
 2}}\left(\frac{\phi^{\prime 2}}{\mathcal H}\right)^\prime\right)k^2 -
 \frac{\phi^{\prime 2}}{2}}{k^2\mathcal
 H^2-\frac{\phi^{\prime 2}}{2}},
\end{align}
where $k^2 = p^2 + 4$.  Note that in the limit of large $p$
($\leftrightarrow$ large $k$) the equation of motion for $\mathcal
R_c^p$ reduces to the standard form, where, defining $z =
a\phi^\prime/\mathcal H$, $A(\eta,p) = z^\prime/z$ and $B(\eta,p) =
p^2$.  Defining horizon crossing as when $A(\eta,p)^2 = B(\eta,p)$,
noting $k^2\ge 4$ and assuming $\dot\phi^2/H^2\ll 1$ and
$1/(a^2H^2)\ll 1$ (which is always the case after curvature domination
in our particular models), we find the condition
\begin{equation}\label{hccond}
 p^2+4 = a^2H^2\left(1+\frac{\ddot\phi}{\dot\phi H}\right)^2 - \left(1+2\frac{\ddot\phi}{\dot\phi H}\right).
\end{equation} 
One can see that if the slow-roll condition $\ddot\phi/(\dot\phi H)\ll1$
is satisfied, then this reduces to the standard horizon-crossing
condition for large $p$.  

\section{\label{models}Toy Models}

Having summarised the key features of open inflation models and the
relevant background and perturbation equations, in this section we revisit
two of the toy models that were introduced and analysed numerically in
\cite{Linde:1999wv}.  

\subsection{Model 1}

The potential of Model 1 takes the form
\begin{equation}\label{m1pot}
 V(\phi) = \frac{1}{2}m^2\phi^2\left(1 + \frac{\alpha^2}{\beta^2 + (\phi-\nu)^2}\right).
\end{equation}
We consider the same parameter values as considered in
\cite{Linde:1999wv}, except we work in reduced Planckian units instead
of Planckian units.  As such, our parameters are given as
\begin{align}\nonumber
 \nu &= 3.5\times\sqrt{8\pi}\\\nonumber
\beta^2 &= 2\alpha^2\\\nonumber
\beta &= 0.1\times \sqrt{8\pi}\\\nonumber
m &= 1.5\times10^{-6}\times \sqrt{8\pi}.
\end{align}
\begin{figure}
\begin{center}
\includegraphics[width = 0.9\columnwidth]{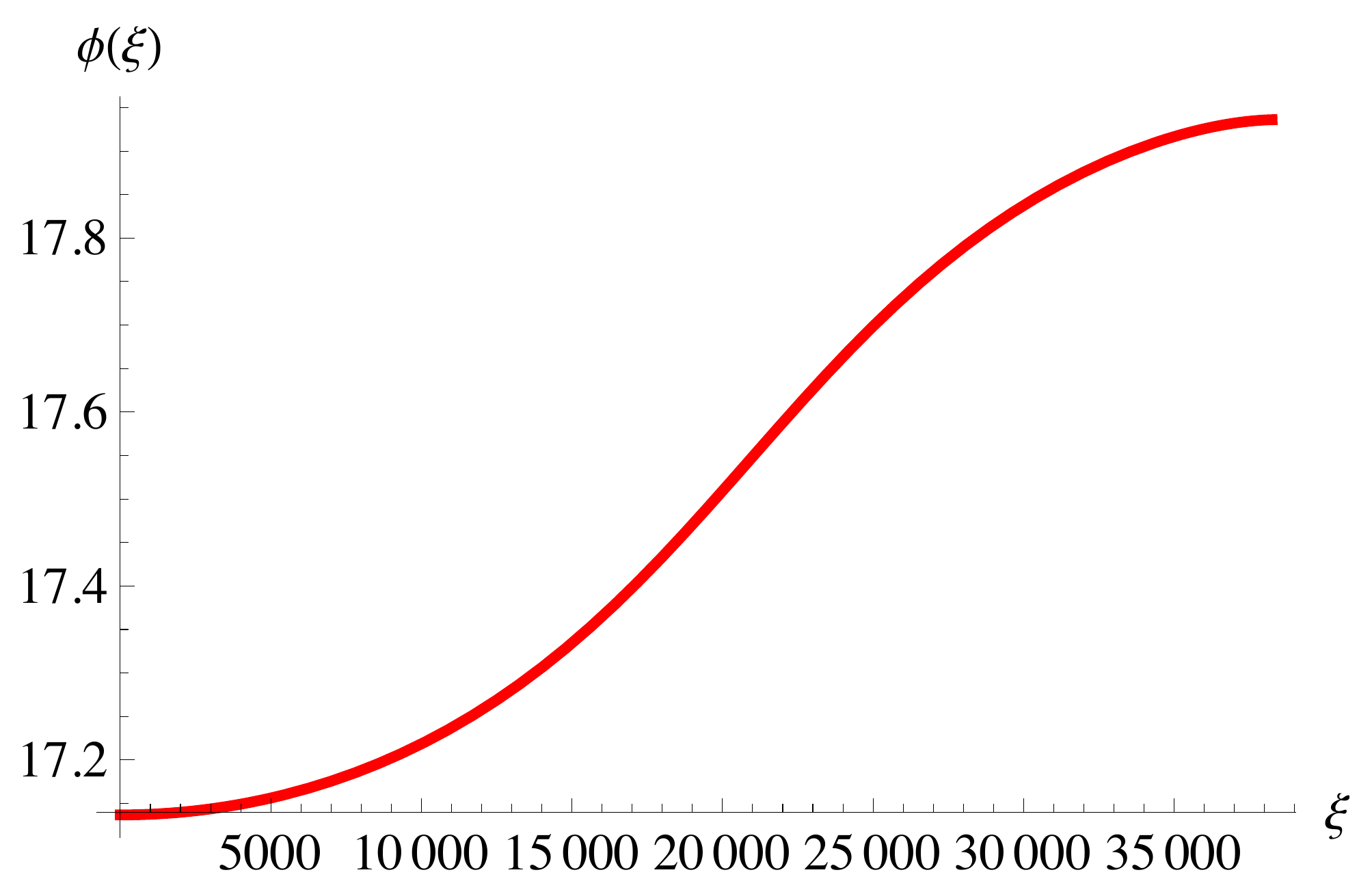}
\includegraphics[width = 0.9\columnwidth]{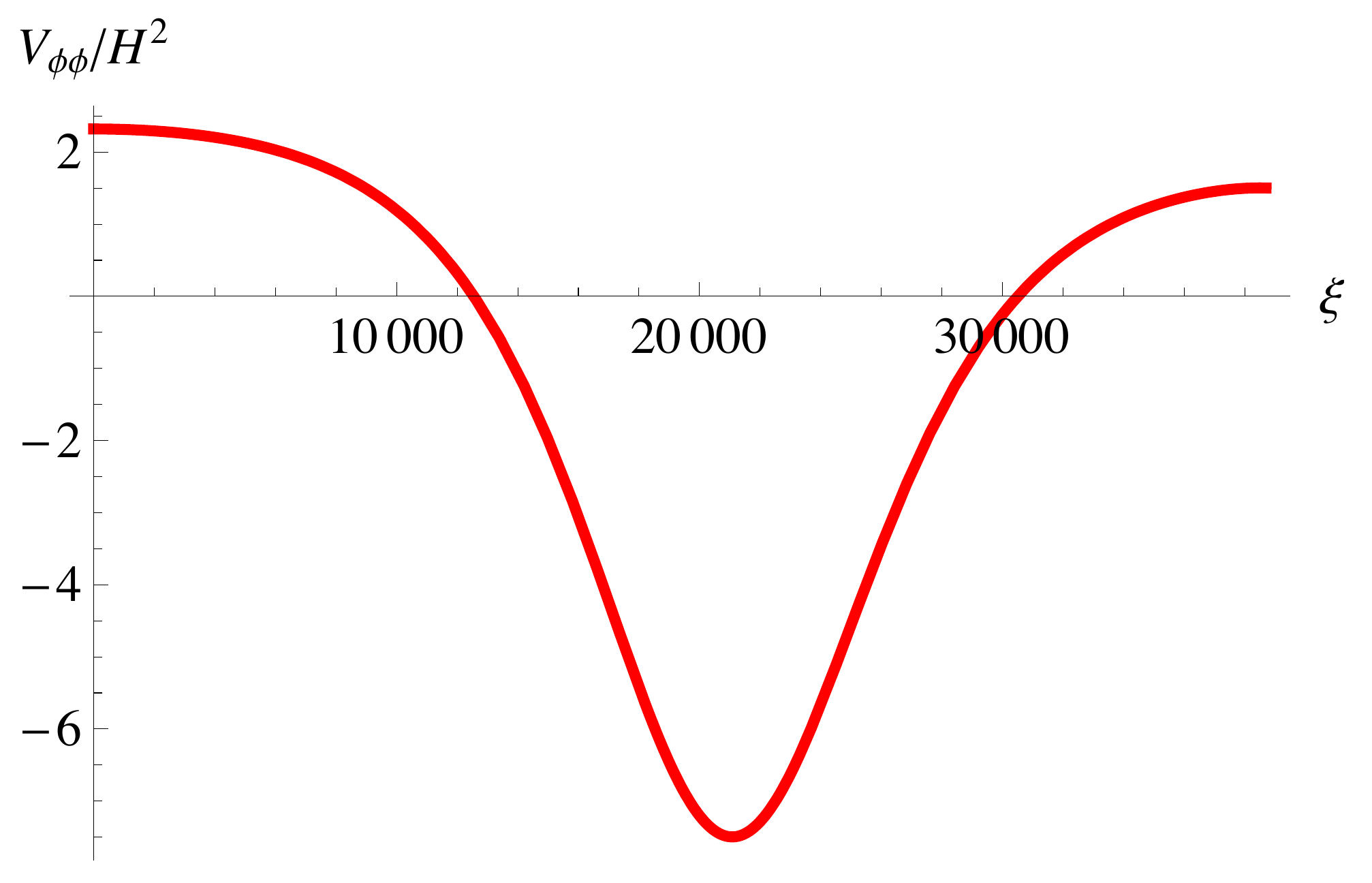}
\caption{\label{phisol}Upper Panel: $\phi$ as a function of $\xi$ for
the CDL instanton solution of Model 1. Lower panel: A plot of
$|V_{\phi\phi}|/H^2$ as a function of $\xi$ for the CDL instanton
solution of Model 1.  We can see that for most of the trajectory the
condition $|V_{\phi\phi}|>H^2$ is satisfied.}
\end{center}
\end{figure}
By solving Eqs.~(\ref{EFried}) and (\ref{Escalar}) numerically, one
is able to find a CDL instanton solution, which is plotted in the upper
panel of figure
\ref{phisol}. The value of the field at the end of the tunnelling is
given as 
\begin{equation}
 \phi_{{\rm exit}} = 17.14.
\end{equation}
In the lower panel of figure \ref{phisol} we plot $|V_{\phi\phi}|/H^2$ along
the CDL instanton trajectory, and we see that $|V_{\phi\phi}|>H^2$ is indeed
satisfied for most of the trajectory.


\begin{figure}[h]
 \centering \includegraphics[width =
\columnwidth]{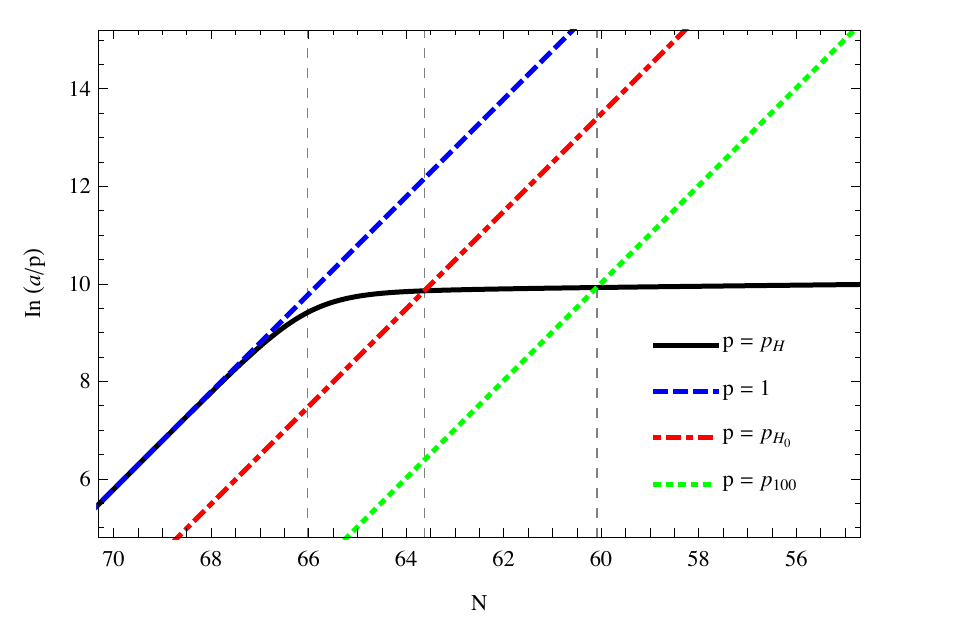} \caption{Evolution of the Hubble
scale as a function of the number of e-foldings before the end of
inflation (solid black curve).  Also plotted are the evolution of the
curvature scale (upper dashed line), the current Hubble scale as determined
assuming that the observational constraint on $\Omega_{{\rm K}}$ is
saturated (middle dot-dashed line) and the scale associated with $l=100$
(bottom dotted line).  The left-most vertical dashed line corresponds to
the time at which the curvature term becomes sub-dominant to the
potential term in the Friedmann equation.  The next two lines, from left
to right, indicate the times at which the current Hubble scale and the
scale associated with $l = 100$ left the Hubble horizon.}\label{m1_he}
\end{figure}

\begin{figure}[h]
\begin{center}
\includegraphics[width = \columnwidth]{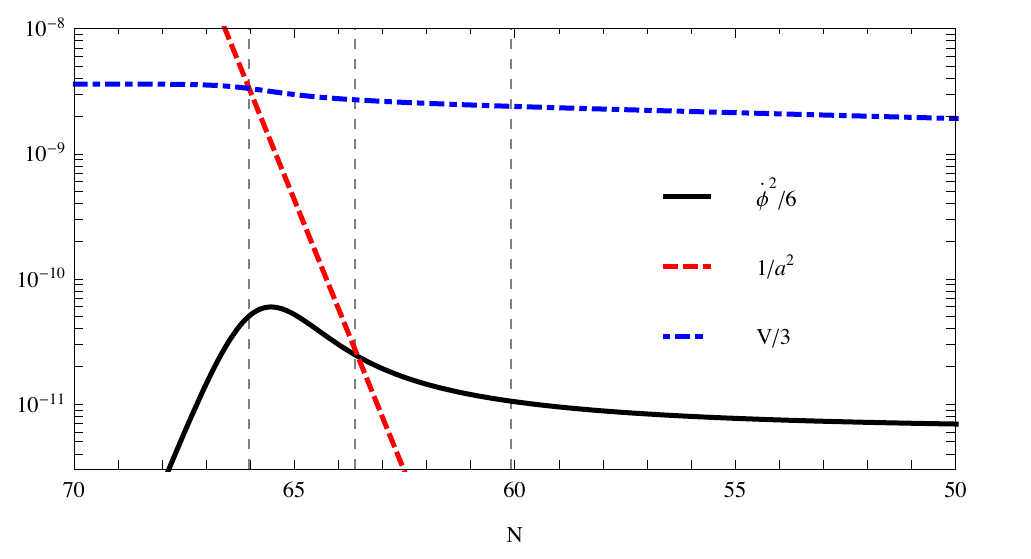}
\caption{\label{m1_fc}Evolution of the kinetic (solid black curve),
 curvature (dashed red curve) and potential (dot-dashed blue curve)
 contributions to $H^2$ as
 a function of the number of e-foldings before the end of inflation.
 The three vertical dashed lines are as in figure \ref{m1_he}.}
\end{center}
\end{figure}

Using the value of $\phi_{{\rm exit}}$ found above, we then solve
the background equations of motion \eqref{bgFe} and \eqref{bgphiem} for
$\phi$ and $a$ after the tunnelling.  In figure \ref{m1_he} we plot the
resulting evolution of the Hubble parameter as a function of the number
of e-foldings before the end of inflation (solid black
curve).\footnote{The end of inflation is taken to be when $\epsilon_V
\equiv (1/2)(V_\phi/V)^2= 1$.}  We also
plot three scales of interest, which are 
\begin{enumerate}
 \item the curvature scale, corresponding to $p = 1$ (blue dashed curve),
\item  the upper limit on the current Hubble scale, as determined by the
       constraint $\Omega_{{\rm K}}\lesssim 0.01$, i.e. $p_{{\rm
       H_0}}= 1/\sqrt{\Omega_{{\rm K}}} \simeq 10$ (red dot-dashed curve), and
\item  the scale that approximately corresponds to the CMB multipole $l =
       100$, as determined assuming that $H_0$ is given by the upper
       limit mentioned above (green dotted curve).  
\end{enumerate}
Finally, the times at which the last two of the aforementioned
scales leave the horizon (as simply determined by $p = aH$) are marked with vertical dashed
lines approximately 64 and 60 e-foldings before the end of inflation,
respectively.  The third, left-most vertical dashed line, located
approximately 66 e-foldings before the end of inflation, corresponds to the
time of potential-curvature equality.  Unless otherwise mentioned, the
vertical dashed lines shown in all following plots will correspond to
these same three events.  As we will see later, it is also important to know
the time at which the curvature becomes sub-dominant to the kinetic component
of $H^2$.  Interestingly, in Model 1 this time almost exactly coincides
with the time at which the scale $p_{H_0}$ left
the horizon, i.e. $\sim 64$ e-foldings before the end of inflation. The evolution of the three
components of $H^2$ are plotted explicitly in figure \ref{m1_fc}.  

We emphasise that the $p_{H_0}$ discussed above corresponds to the current Hubble scale {\it as
determined assuming that the observational constraint on $\Omega_{{\rm K}}$ is
saturated}.  For our choice of model parameters, we see that this
scale would leave the horizon approximately $64$ e-foldings before the
end of inflation.  However, this would seem a little too early, as
standard arguments dictate that the current horizon scale left the horizon
$\sim 60$ e-foldings before the end of inflation \cite{Liddle:2003as}.  

\begin{figure}[h]
 \centering
\includegraphics[width = \columnwidth]{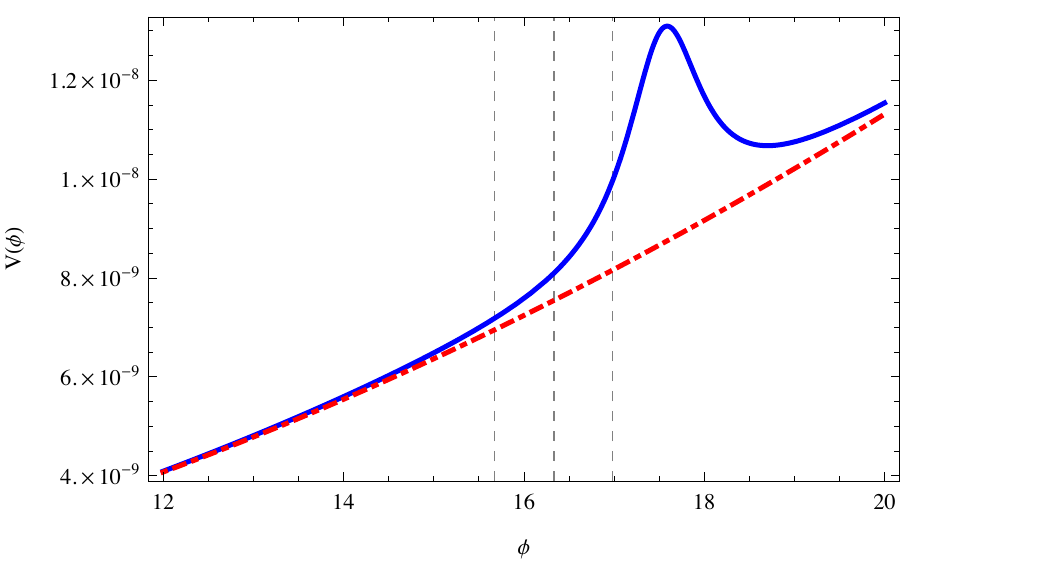}
\caption{Plot of the potential in Model 1 of \cite{Linde:1999wv}, with the parameter choice shown in the main text.  The
 three vertical lines are as in figure \ref{m1_he}, except their order from
 left to right is reversed.}\label{mod1pot}
\end{figure}

In order to help relate features in the background dynamics and power
spectrum to the underlying potential, in figure \ref{mod1pot} we
indicate the location of the scalar field on the potential at the three
times discussed above.  Note that the order of the vertical dashed lines
is the reverse of that in other plots.  Looking at the form of the
potential \eqref{m1pot}, we see that it naturally lends itself to being
decomposed into a fiducial $m^2\phi^2$ potential and a correction to
this around the tunnelling barrier.  As such, in figure \ref{mod1pot} we
also plot the fiducial $m^2\phi^2$ potential.  We see that the
steepening of the potential relative to $m^2\phi^2$ remains
non-negligible until after the scale $p_{100}$ leaves the horizon.  As
such, we expect that observable scales will indeed be subject to
suppression in this model.

\begin{figure}[h]
 \centering \includegraphics[width =
0.9\columnwidth]{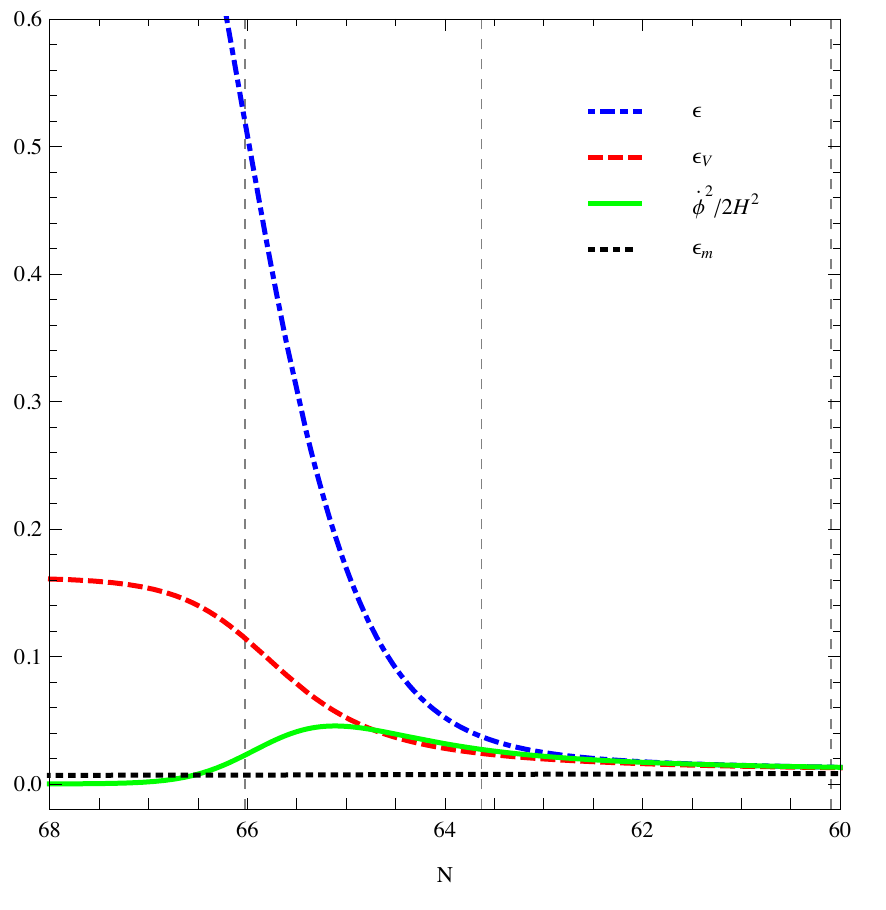} \includegraphics[width =
0.9\columnwidth]{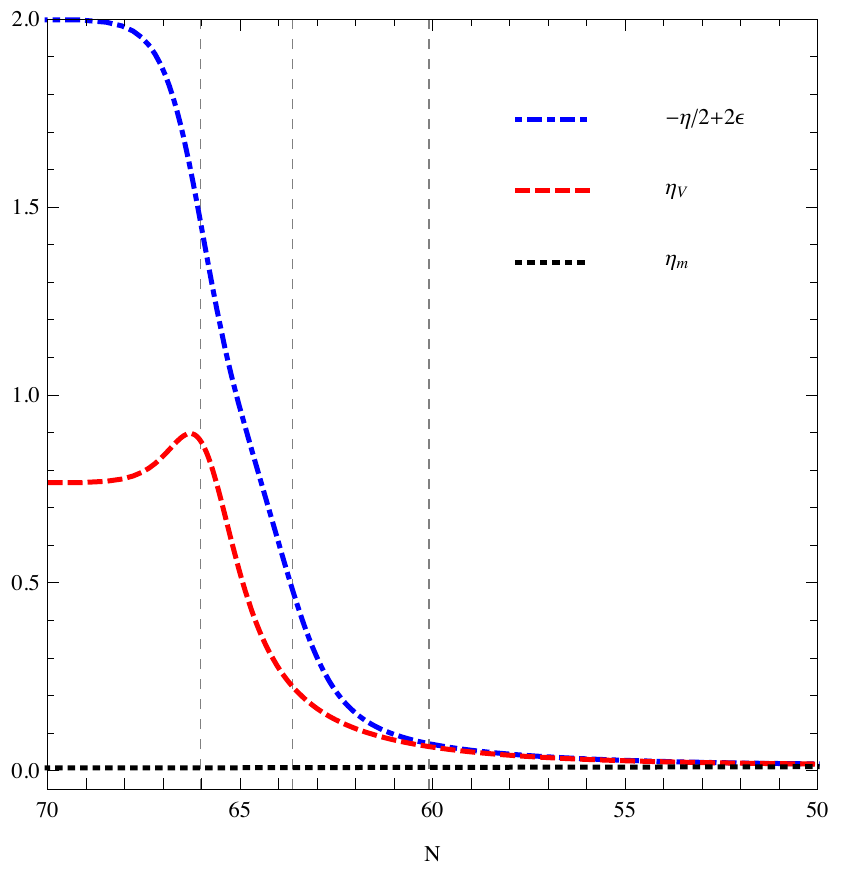} \caption{Upper panel: Evolution of the four
slow-roll quantities $\epsilon$ (blue dot-dashed curve), $\epsilon_V$
(red dashed curve), $\epsilon_m$ (black dotted curve) and  $\dot\phi^2/2H^2$ (solid green curve) as a
function of the number of e-foldings before the end of inflation.  Lower panel:
Evolution of the three slow-roll quantities $\tilde\eta \equiv
-\eta/2+2\epsilon$ (blue dot-dashed curve), $\eta_V$ (red dashed curve)
and $\eta_m$ (black dotted curve) as a function of the number of
e-foldings before the end of inflation.  In both panels the
three vertical lines are as in figure \ref{m1_he}.}\label{m1_sr_param_ev}
\end{figure}

In the upper panel of figure \ref{m1_sr_param_ev} we compare the evolution of the three slow-roll
parameters $\epsilon\equiv -\dot H/H^2$, $\epsilon_V\equiv
(V_\phi/V)^2/2$ and $\epsilon_m = 2/\phi^2$ with the evolution of
$\dot\phi^2/2H^2$.  As we can see from \eqref{scalar_mag}, the quantity
that we are interested in evaluating is $\mathcal P_{\mathcal R}\propto
H^4/\dot\phi^2$.  In the case of a flat universe, this can be
re-expressed in terms of $\epsilon$ as $\mathcal P_{\mathcal R}\propto
H^2/\epsilon$, which in turn, under the slow-roll approximation can be
re-expressed as $\mathcal P_{\mathcal R}\propto H^2/\epsilon_V$.  In the
case of open inflation, however, we have $\dot H = -\dot\phi^2/2 -
1/a^2$, meaning that we can no longer directly replace $\dot\phi^2$ with
$\epsilon$.  Moreover, as the slow-roll condition may be violated in the
early stages of evolution after tunnelling, we are also not necessarily able to
make the second approximation $\epsilon \simeq \epsilon_V$.  Of course,
at late times we expect the curvature to become negligible and for
slow-roll inflation to take place, so it is interesting to explicitly
see at what stage the equivalence of the three quantities is recovered.
As we can see from figure \ref{m1_sr_param_ev}, the equivalence is
recovered approximately $63$ e-foldings before the end of
inflation.  This time is shortly after that at which the curvature term
becomes negligible in comparison to both the potential {\it and} kinetic
terms in the Friedmann equation.
This is exactly as we would expect in light of the contribution of
$1/a^2$ to $\dot H$.  The fact that $\epsilon$ and
$\epsilon_V$ also become approximately equal at this time tells us that
the potential must already be relatively flat and the slow-roll
approximation is a good one.
Another point to note from this figure is the difference between
$\epsilon$ and $\dot\phi^2$ at early times.  As a result of the initial
conditions imposed by the CDL instanton and the initial curvature domination,
$\epsilon$ asymptotes to unity whilst $\dot\phi$ vanishes.  This is
evidently very important in determining the asymptotic behaviour of
$\mathcal P_{\mathcal R}$, as we will see shortly.  Finally the
$\epsilon_m$ that we have plotted corresponds to the slow-roll parameter
associated with the fiducial $m^2\phi^2$ potential.  We can see that
even 60 e-foldings before the end of inflation there is a
non-negligible difference between $\epsilon_m$ and $\epsilon_V$.

To leading order in the slow-roll approximation, the tilt of the power
spectrum is given as
\begin{equation}
 n_s-1 = -2\epsilon-\eta,
\end{equation}
where $\eta \equiv \dot\epsilon/(H\epsilon)$.  Using
the fact that $\eta_V\equiv V_{\phi\phi}/V \simeq -\eta/2 + 2\epsilon$, this can be written
in the perhaps more familiar form
\begin{equation}
 n_s -1 = -6\epsilon_V + 2\eta_V.
\end{equation}
We know that the initial conditions prescribed by the CDL instanton
dictate that $\eta_V$ be large, which will tend to give a blue-tilted
spectrum.  Given that the spectrum must be red-tilted on small scales,
we are therefore interested in following the evolution of $\eta$ and in
determining when the transition from a blue- to a red-tilted spectrum
takes place.  As such, in the lower panel of figure \ref{m1_sr_param_ev}
we plot the evolution of $\tilde\eta \equiv -\eta/2 + 2\epsilon$,
$\eta_V$ and $\eta_m\equiv 2/\phi^2$, where the last quantity
corresponds to the equivalent slow-roll parameter associated with the
fiducial $m^2\phi^2$ potential.  Similar to the case with $\epsilon$ and
$\epsilon_V$, we note the different behaviour of $\tilde\eta$ and
$\eta_V$ at early times.  This is again due to the initial conditions
imposed by the CDL instanton and the domination of curvature, which give
$\eta = 0$ and $\epsilon = 1$ as $t\rightarrow 0$.  Comparing with the upper panel of
figure \ref{m1_sr_param_ev}, we also note that the time at which the three
different definitions of $\eta$ become equivalent is somewhat later than
the corresponding time for $\epsilon$.

\begin{figure}
\begin{center}
\includegraphics[width = \columnwidth]{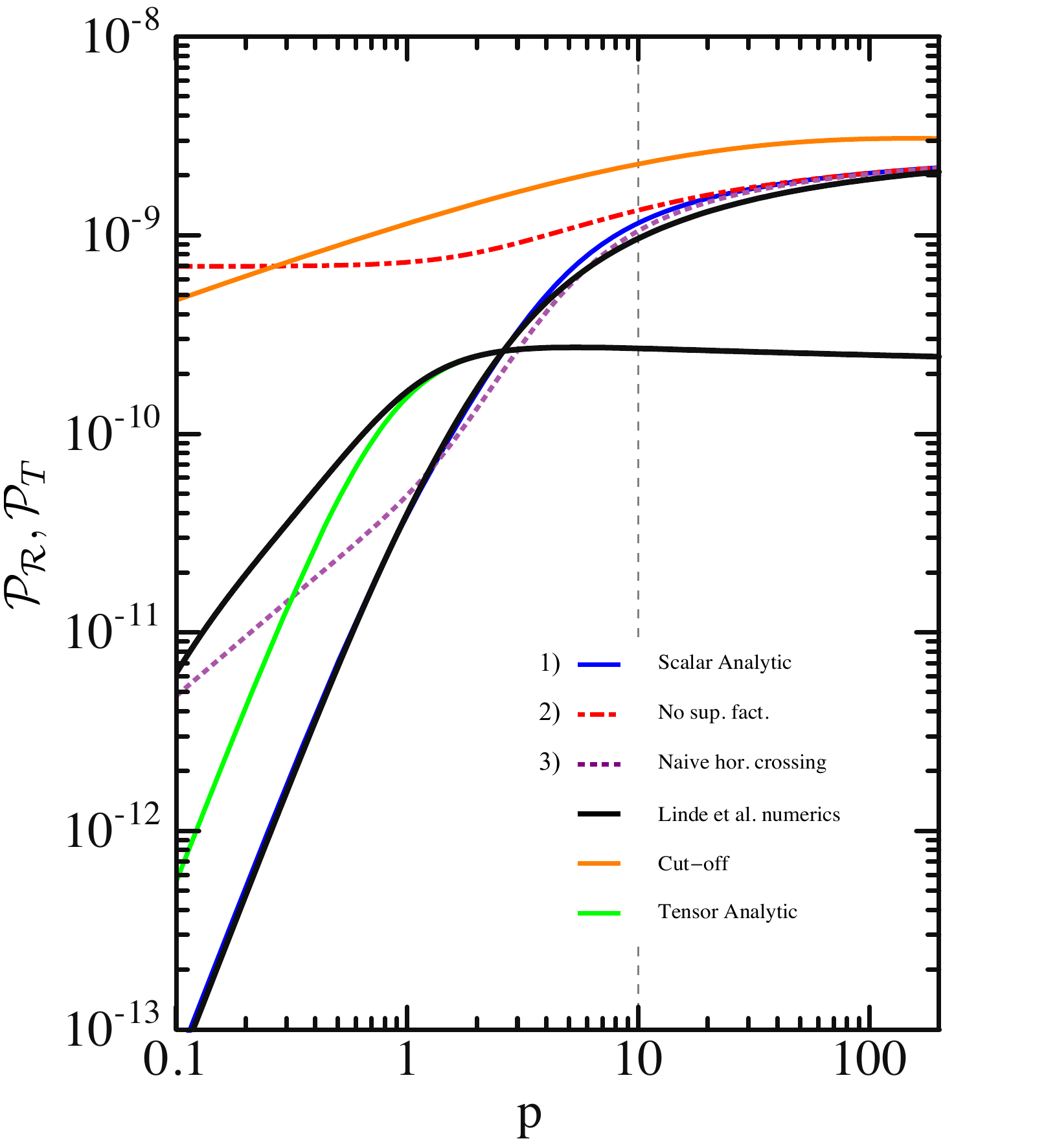}
\caption{\label{m1_spectra}The scalar and tensor power spectra as
functions of the comoving scale $p$.  In the case of the scalar spectrum we
plot three variants of the analytic expression \eqref{scalar_mag},
namely 1) the full expression (solid blue curve), 2) \eqref{scalar_mag}
without the $p$-dependent suppression factor that distinguishes
\eqref{scalar_mag} from the standard flat-universe expression (red
dot-dashed curve) and 3) \eqref{scalar_mag} taking $t_{\mathcal R, p}$
as the time when $a^2H^2 = p^2+1$ instead of when \eqref{hccond} is
satisfied (purple dotted curve).  We also plot the numerical solution
for $\mathcal R^p_c$ from \cite{Linde:1999wv} (upper solid black curve).
The fitted curves use $c_1 = 4$.  Finally, for comparison, we also plot
the form of the spectrum associated with the cut-off model considered in
\cite{Ade:2013uln} (solid orange curve).  In the case of the tensor
spectrum we simply plot the full expression \eqref{tensor_mag} with $c_2
= 1$ (solid green curve) and the numerical solution from
\cite{Linde:1999wv} (lower solid black curve). The vertical dashed line
at $p=10$ corresponds to the current Hubble scale if we assume
that observational constraints on $\Omega_{{\rm K}}$ are saturated.}
\end{center}
\end{figure}

In figure \ref{m1_spectra} we plot the scalar and tensor power spectra
associated with Model 1 as functions of the comoving scale $p$.
In the case of the scalar spectrum we plot four different curves.  Three
of these correspond to different variants of the analytic expression
\eqref{scalar_mag}, namely 1) the full expression \eqref{scalar_mag}
(blue solid curve) 2) \eqref{scalar_mag} without the $p$-dependent
suppression factor that distinguishes \eqref{scalar_mag} from the
standard flat-universe expression (red dot-dashed curve) and 3)
\eqref{scalar_mag} taking $t_{\mathcal R, p}$ as the time when $a^2H^2 =
p^2+1$ instead of when the condition \eqref{hccond} is satisfied (purple
dotted curve).  The upper solid black curve then corresponds to the
numerical solution for $\mathcal R^p_c$ found in \cite{Linde:1999wv}.
We take $c_1 = 4$ in \eqref{scalar_mag}, which we find to give the best
fit to the numerical results for small $p$.  Qualitatively, we see that
the full analytic result (solid blue curve) matches the numerics well,
especially for small and large $p$.  At intermediate scales, around $p
\simeq 10$, however, it tends to underestimate the amount of
suppression, with the discrepancy being on the order of 10\%. If we
assume that the observational bound on $\Omega_{{\rm K}}$ is saturated,
this scale coincides exactly with the current Horizon scale.  As such,
if we are interested in quantitatively constraining the suppression from
open inflation, we see that numerical calculations are necessary.

From curves 2) and 3) we are able to appreciate the importance of
corrections to the standard flat-universe expression for $\mathcal
P_{\mathcal R}$ for small $p$.  The fact that curve 2) always
underestimates the suppression of the scalar spectrum highlights the fact that
the fast-rolling of the inflaton is not the only source of suppression,
especially on very large scales.  The additional suppression reflects
the system's memory of the tunnelling process preceding inflation
\cite{Linde:1999wv}.  We also see that curve 3) underestimates the
suppression for $p\lesssim 1$.  This reflects the fact that large-scale
modes freeze out at later times than predicted by the standard
horizon-crossing condition, leading to a suppression in their amplitude.
As expected, all curves converge for large $p$, where the effects of the
tunnelling become negligible.

In addition to the four curves discussed above, for comparison we also
plot the spectrum associated with the so-called cut-off model that has
been considered in the literature, see
e.g. \cite{Contaldi:2003zv,Cline:2003ve}.  The model -- corresponding to
the solid orange curve in figure \ref{m1_spectra} -- is a
phenomenological example of a spectrum with suppression on large scales,
and takes the form
\begin{equation}
\mathcal P_{{\rm co}} = A_s\left(\frac{p}{p_\ast}\right)^{n_s-1}\left[1-\exp\left\{-\left(\frac{p}{p_c}\right)^{\lambda_c}\right\}\right].
\end{equation}
The {\it Planck} team found that such a form for the spectrum was preferred
over power-law $\Lambda$CDM with $2\Delta\ln \mathcal{L}_{{\rm max}}=2.9$
\cite{Ade:2013uln}.  The corresponding best-fit parameters were $\ln
(p_c/(a_0\, {\rm Mpc}^{-1})) = -8.493$ and $\lambda_c = 0.474$, where
the pivot scale $p_\ast/a_0 = 0.05\,{\rm Mpc^{-1}}$ was used.  Whilst it
is not our intention that our models quantitatively fit the data well,
it is nevertheless interesting to qualitatively compare our spectra with
this cut-off model.  For the orange curve in figure \ref{m1_spectra} we have
used {\it Planck}'s best-fit values for $p_c/a_0$ and $\lambda_c$ and
further assumed that $\Omega_{{\rm K}} = 0.01$ and $n_s = 0.96$.  The amplitude $A_s$ is adjusted so that
our spectrum and the phenomenological model agree at $p = 10^5$, as both
spectra have relaxed to the standard power-law form on these scales.

In comparing Model 1 with the cut-off spectrum, the main feature we note
is that Model 1 gives substantially more suppression.  This
feature is favourable in light of the fact that {\it Planck}'s best-fit
spectrum was found assuming no tensor modes.  In order to be able to
accommodate a tensor contribution to large-scale CMB
temperature fluctuations (as suggested by BICEP2), we expect that a
larger suppression of the scalar spectrum will be required.  The
discrepancy between the two models is, however, rather substantial, even
at $p \simeq 100$.  Better agreement could perhaps be achieved by
correcting our assumption that $\Omega_{{\rm K}} = 0.01$.  As such, in Model 1 we
may find that the best-fit value for $\Omega_{{\rm K}}$ is unobservably
small.

For reference, we also plot the tensor spectrum, but this time only
include the full analytic result given by \eqref{tensor_mag} with $c_2 =
1$ (solid green curve) and the numerical results from
\cite{Linde:1999wv} (lower solid black curve).  It is clear to see that
the onset of suppression occurs at much smaller $p$ in the case of the
tensor spectrum, i.e. on scales larger than the current Hubble horizon.
In contrast with the scalar spectrum, the tensor spectrum is not
affected by the fast rolling of the inflaton.  As such, the suppression
on large scales is purely a result of the pre-inflationary tunnelling
and the presence of the bubble wall.  See \cite{Linde:1999wv} for a more
detailed discussion.

\begin{figure}
\begin{center}
\includegraphics[width = \columnwidth]{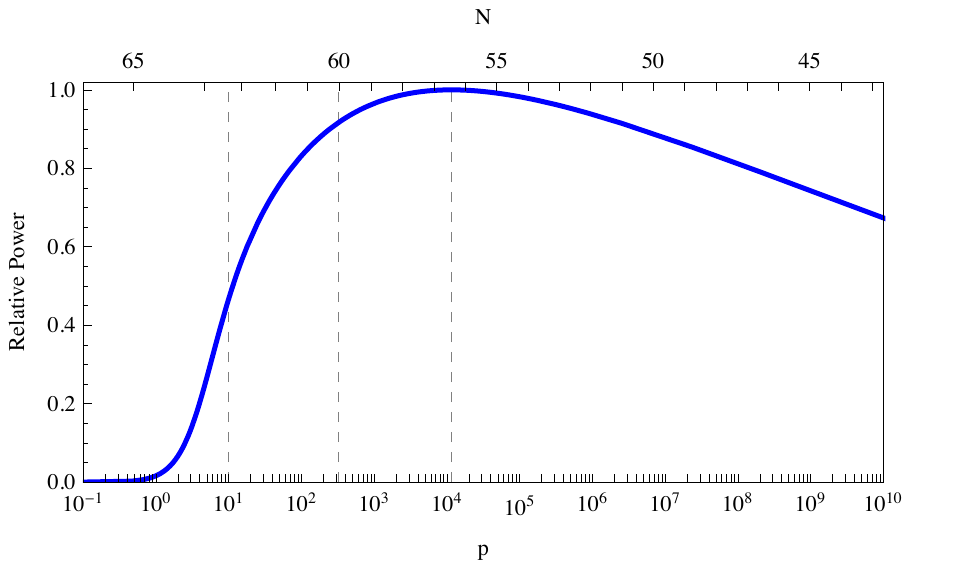}
\caption{\label{m1_relpow}The magnitude of the scalar power relative to
 the magnitude at the scale of
 transition between a blue- and red-tilted spectrum.  In this plot the
 vertical dashed lines, from left to right, correspond, respectively, to 1) the current
 Horizon scale $p_{H_0}$ as determined assuming that $\Omega_{{\rm K}}
 =0.01$ 2) the scale associated with $l = 100$$, p_{100}$, and 3) the
 scale at which the spectrum transitions from being blue- to red-tilted,
 $p_{{\rm red}}$.  On the upper
 horizontal axis we also give the number of e-foldings before the end of
 inflation that each given scale $p$ leaves the
 horizon.}
\end{center}
\end{figure}

Finally, in figure \ref{m1_relpow} we plot the magnitude of the scalar
power relative to that at the scale at which the spectrum transitions
from being blue- to red-tilted.  This transition occurs approximately 56
e-foldings before the end of inflation, when the scale $p_{{\rm
red}}\simeq 10^4$ left the horizon.  Given that
potential-curvature equality occurs approximately 66 e-foldings before
the end of inflation, we find that we have roughly 10 e-foldings of the
``fast roll'' phase before the spectrum becomes red-tilted and standard
slow-roll is achieved.     

Even in the best-case scenario, it is only possible to detect a
$\Omega_{{\rm K}}\gtrsim 10^{-4}$, which would correspond to $p_{H_0} =
10^2$.  As such, we see that in Model 1 we have the possibility that
suppression of the scalar spectrum is observed on large scales whilst
$\Omega_{{\rm K}}$ remains undetectable.  Indeed, if we were to make the usual
assumption that $p_{H_0}$ leaves the horizon $\sim 60$ e-foldings
before the end of inflation, this would correspond to $p_{H_0}\sim
3\times 10^2$, giving an unobservable $\Omega_{{\rm K}}\sim 1\times 10^{-5}$.
However, in such a scenario the scale $p_{{\rm red}}$ would coincide
with $l \sim 100$, and suppression on the order of 10\% would be
observed at scales corresponding to the current Hubble horizon.

\subsection{Model 2}

The second model considered in \cite{Linde:1999wv} has a potential of
the form  
\begin{equation}\label{pot2}
 V(\phi) = \frac{m^2}{2}\left(\phi^2 - B^2\frac{\sinh\left[A(\phi-\nu)\right]}{\cosh^2\left[A(\phi-\nu)\right]} \right),
\end{equation}
with parameters chosen as 
\begin{align}\nonumber
 m &= 10^{-6}\times\sqrt{8\pi}\\\nonumber
A &= 20/\sqrt{8\pi}\\\nonumber
B &= 4\times \sqrt{8\pi}\\\nonumber
\nu &= 3.5\times\sqrt{8\pi}.
\end{align}

\begin{figure}
\begin{center}
\includegraphics[width = 0.9\columnwidth]{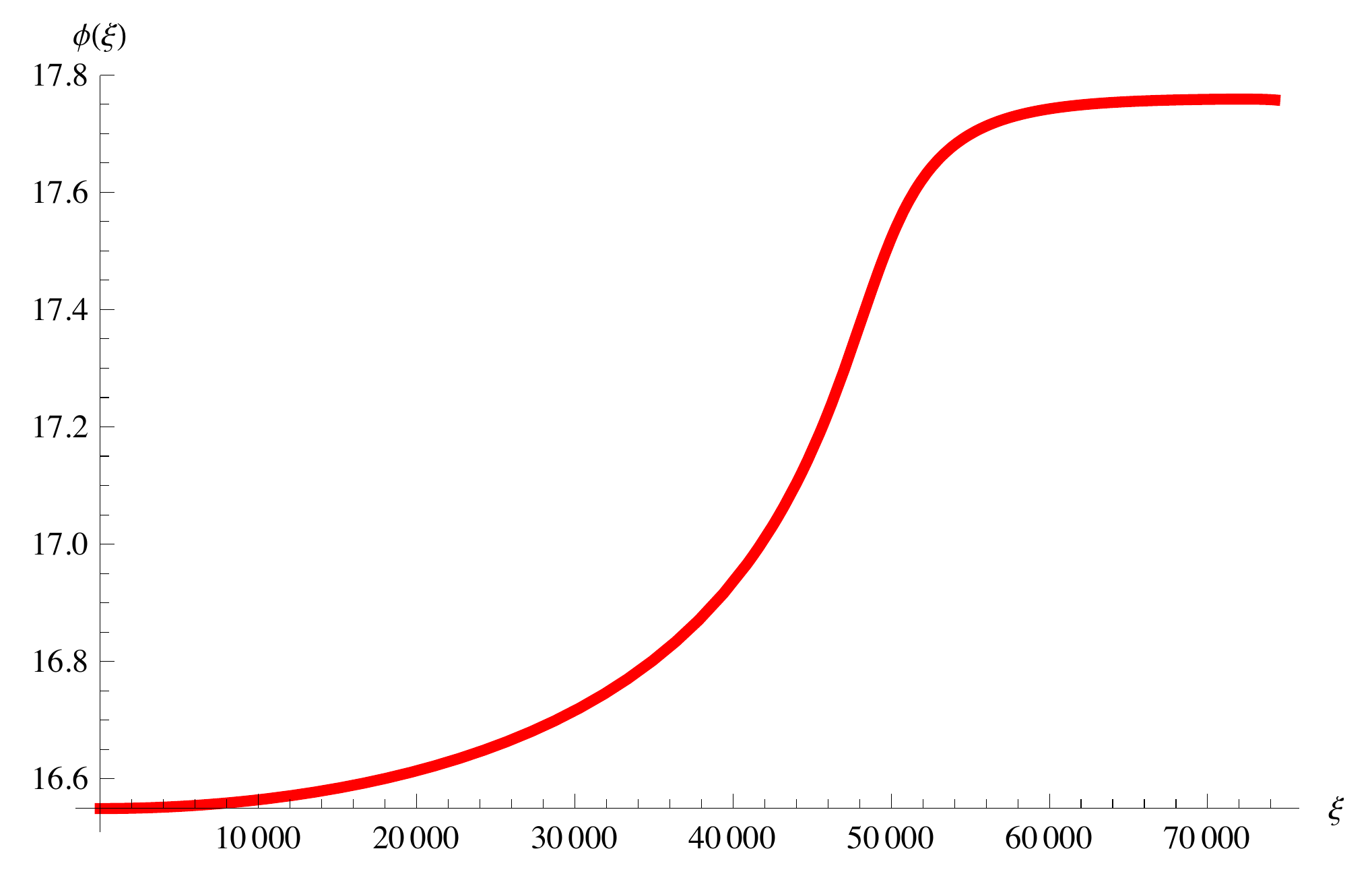}
\includegraphics[width = 0.9\columnwidth]{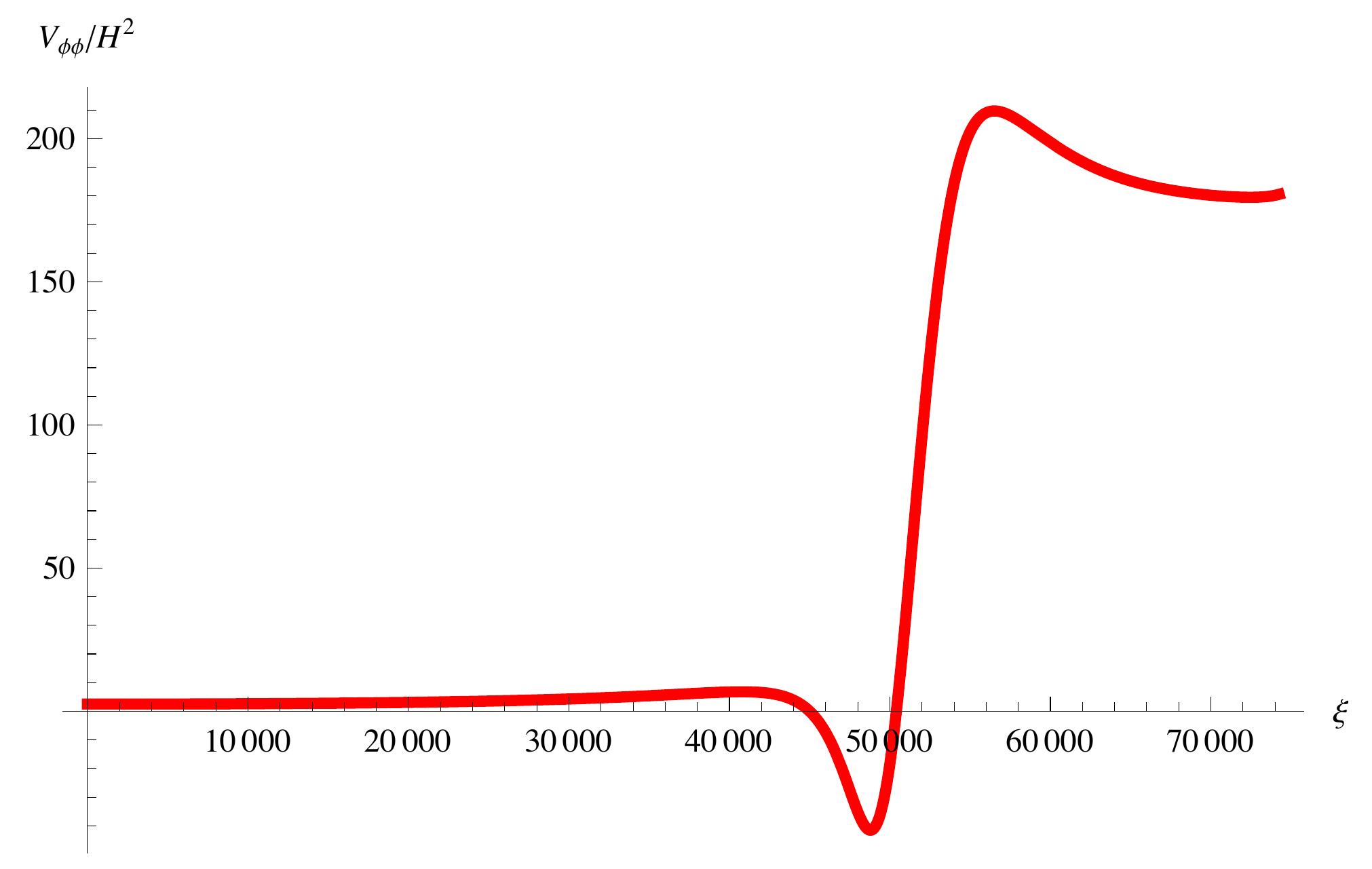}
\caption{\label{CDL2}Upper Panel: $\phi$ as a function of $\xi$ for
the CDL instanton solution of Model 2. Lower panel: A plot of
$|V_{\phi\phi}|/H^2$ as a function of $\xi$ for the CDL instanton
solution of Model 2.  We can see that for most of the trajectory the
condition $|V_{\phi\phi}|>H^2$ is satisfied.}
\end{center}
\end{figure}

The CDL instanton solution and the evolution of $V_{\phi\phi}/H^2$
through the tunnelling process are shown in figure \ref{CDL2}.  From
the upper plot we are able to obtain $\phi_{{\rm exit}}=16.55$,
and from the lower plot we see that the condition $|V_{\phi\phi}|>H^2$ is
indeed satisfied for most of the trajectory.

Qualitatively, the features of Model 2 are very similar to Model 1,
except that the dependence of the potential on hyperbolic trigonometric
functions makes deviations from the fiducial $m^2\phi^2$ potential
around the tunnelling barrier sharper and more localised.  Given the
similarities, we refrain from reproducing the full set of plots given in
the case of Model 1, focusing only on those we feel highlight the
differences between the two models.

\begin{figure}[h]
 \centering
\includegraphics[width = \columnwidth]{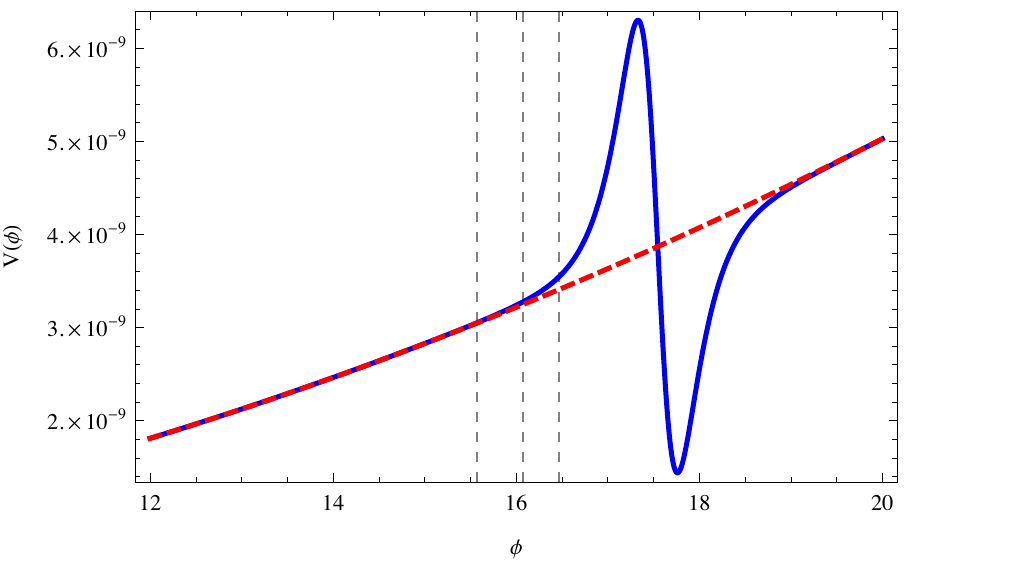}
\caption{Plot of the potential in Model 2 of \cite{Linde:1999wv} with the parameter choice shown in the main text.  The
 three vertical lines are as in figure \ref{mod1pot}.}\label{m2_pot}
\end{figure}

In order to highlight the sharpness of the barrier feature,
in figure \ref{m2_pot} we plot Model 2's equivalent of figure
\ref{mod1pot}.  Namely, we plot the potential, the corresponding fiducial
$m^2\phi^2$ potential and the position of the field at the time of
potential-curvature equality, the time that $p_{H_0}$ left the
horizon and the time that $p_{100}$ left the horizon.  We see that
the potential relaxes to the fiducial one at around the time that $
p_{H_0}$ left the horizon, which is much sooner than in the case of
Model 1.  As such, we expect that a much smaller range
of scales will be subject to suppression.

\begin{figure}
\begin{center}
\includegraphics[width = \columnwidth]{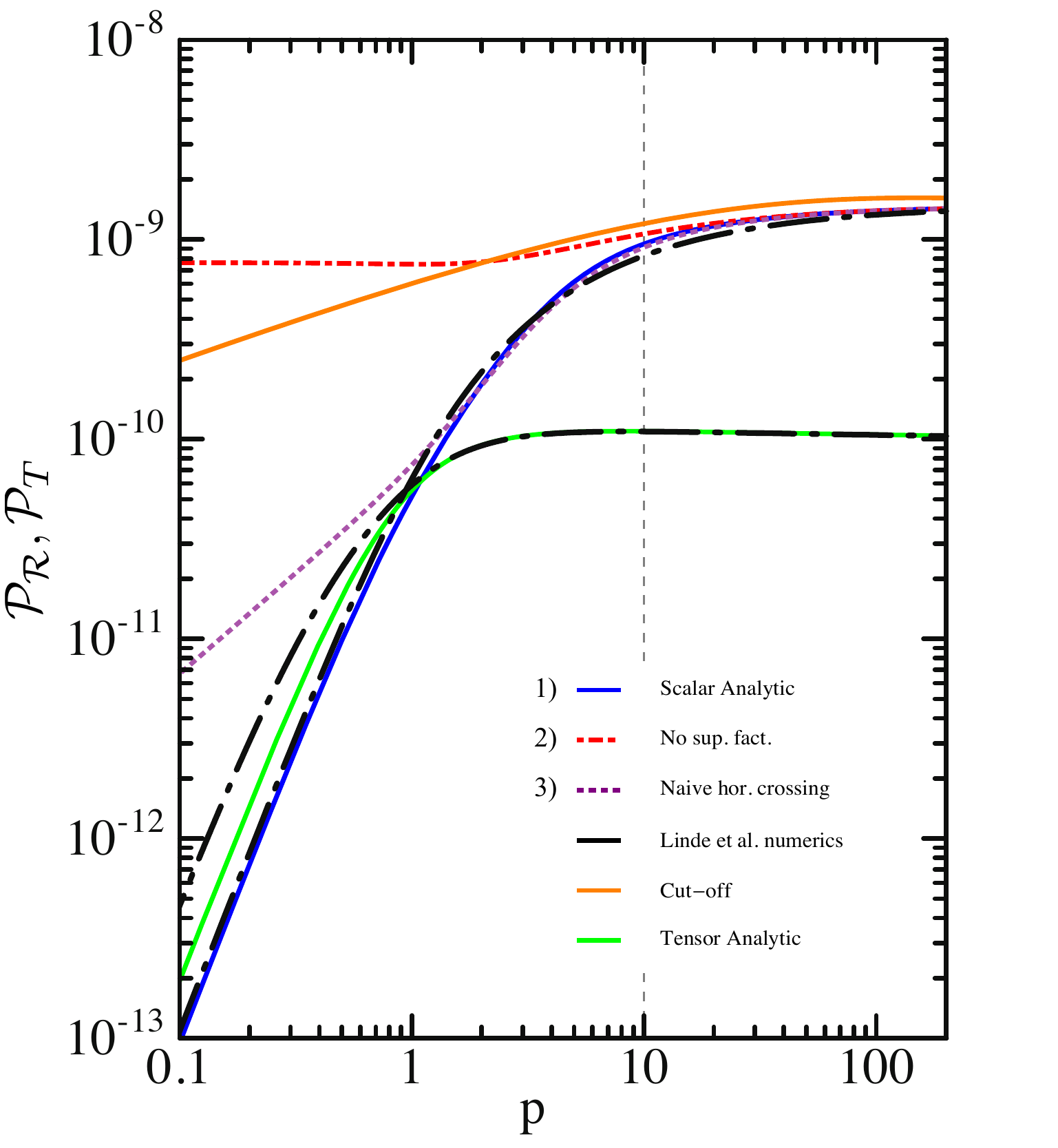}
\caption{\label{m2_spectra}The scalar and tensor power spectra as a function of
the comoving scale $p$ for Model 2.  The set of curves plotted is the same as in figure \ref{m1_spectra} for Model 1, except that solid black curves are now replaced with black dot-dashed curves.}
\end{center}
\end{figure}

In figure \ref{m2_spectra} we plot the scalar and tensor power spectra
as a function of $p$, with the same set of curves included here as in
figure \ref{m1_spectra}.  Regarding the scalar spectrum, in the case of
Model 2 we find that using $c_1 = 3.5$ in \eqref{scalar_mag} gives a
better fit to the numerical results of Linde {\it et al.} for small $\rm
p$.  Once again we find that at intermediate scales, i.e. $p \simeq 10$,
the analytic fit differs from the numerical results by an amount on the
order of 10\%.  We also see that neglecting the $p$-dependent
suppression factor in \eqref{scalar_mag} and using only the naive
horizon-crossing condition to determine $t_{\mathcal R,p}$ have very
similar effects as in the case of Model 1.  The cut-off spectrum
represented by the solid orange curve is constructed in exactly the same
way as in the case of Model 1.  As with Model 1, we find that the
spectrum of Model 2 is much more suppressed on large scales compared to
the cut-off model.  A tensor contribution to CMB temperature
fluctuations could therefore be accommodated in the case that the
results of BICEP2 are confirmed.  In comparison with Model 1, Model 2 is
in much better agreement with the cut-off spectrum on smaller scales.
Regarding the tensor spectrum, we once again use $c_2 = 1$ in our fit,
and again we can see that suppression of the tensor spectrum only
becomes active at much lower $p$ than in the case of the scalar
spectrum.

\begin{figure}
\begin{center}
\includegraphics[width = \columnwidth]{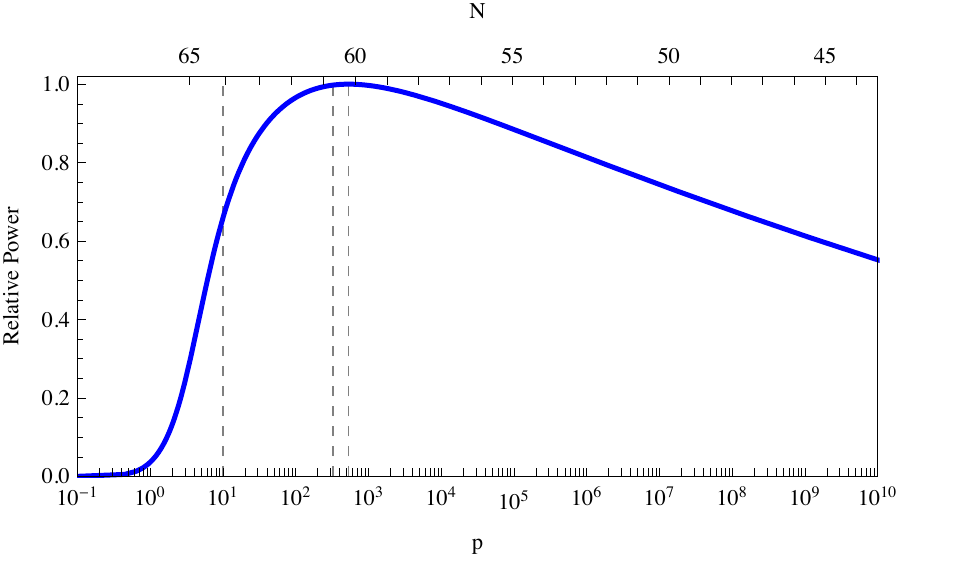}
\caption{\label{m2_relpow}The magnitude of the scalar power relative to
 the magnitude at the scale of
 transition between a blue- and a red-tilted spectrum for Model 2.  In this plot the
 vertical dashed lines, from left to right, correspond respectively to 1) the current
 horizon scale $p_{H_0}$ as determined assuming that $\Omega_{{\rm K}}
 =0.01$ 2) the scale associated with $l = 100$$, p_{100}$, and 3) the
 scale at which the spectrum transitions from being blue- to red-tilted,
 $p_{{\rm red}}$.  On the upper
 horizontal axis we also give the number of e-foldings before the end of
 inflation that each given scale $p$ leaves the
 horizon.}
\end{center}
\end{figure}

Finally, in figure \ref{m2_relpow} we plot the magnitude of the scalar
power spectrum relative to its value at the scale of the transition
between a blue- and a red-tilted spectrum.  It is in this plot that the
differences between Models 1 and 2 become most apparent.  Comparing with
figure \ref{m1_relpow} we see that the transition from a blue- to
a red-tilted spectrum occurs a few e-foldings earlier in Model 2,
approximately 60 e-foldings before the end of inflation.  Given that
potential-curvature equality occurs approximately 66 e-foldings before
the end of inflation, this means that we only have around 6 e-foldings
of the ``fast-roll'' phase.  Correspondingly, a suppression of the scalar
power spectrum is observable over a much smaller range of scales, with
the smallest affected scales being quite close to $p_{100}$ in the case
that $\Omega_{{\rm K}} = 0.01$.  Nevertheless, the model is still able
to give rise to a suppression on observationally relevant scales whilst
also satisfying constraints on $\Omega_{{\rm K}}$.  Unlike Model 1,
however, we see that if $\Omega_{{\rm K}}$ is of non-detectable magnitude,
i.e. $\Omega_{{\rm K}}\lesssim 10^{-4}\rightarrow p_{H_0}\gtrsim
10^2$, then this model likely gives no suppression on observable scales.
Similarly, if we were to make the standard assumption that the current
Hubble scale left the horizon $\sim 60$ e-foldings before the end of
inflation, then we see that there would be no suppression on observable
scales.

\section{\label{general}General constraints}

As discussed in the introduction, our motivation for considering open
inflation models is that they naturally give rise to a fast-roll phase
between tunnelling from the false vacuum and the onset of slow-roll
inflation, which in turn leads to a suppression of the scalar power
spectrum for modes leaving the horizon during this period.  However, in
order for this suppression to be observable, we
require that the total number of e-foldings of inflation be relatively small.  This
in turn means that the curvature of space may not get diluted enough to
satisfy current observational constraints on $\Omega_{{\rm K}}$.  In
this section we summarise how observational
constraints on these two competing effects might be used to constrain
models of open inflation.

\begin{figure}[t]
 \centering \includegraphics[clip=false, trim= 0 20 0 20, width =
\columnwidth]{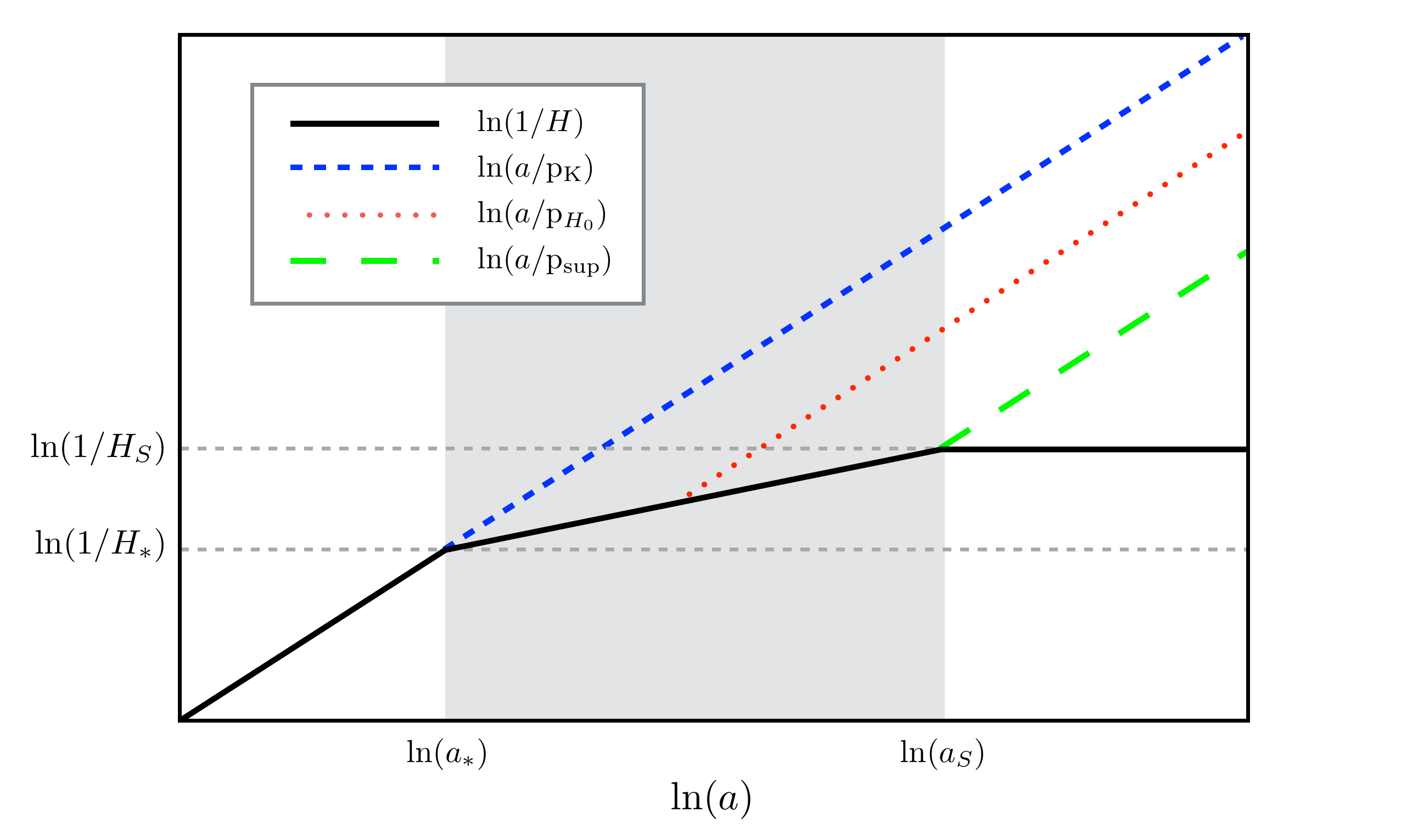} \caption{Schematic diagram showing the
 evolution of $\ln(1/H)$ (solid black curve) through the
three stages of post-tunnelling evolution referred to in the main text.
Initially the curvature dominates,
giving $1/H\propto a$.  In the second stage the potential now dominates
but the slope of the potential is too steep to allow for slow-roll
inflation, so the Hubble rate continues to evolve, but less rapidly.
Once the potential flattens out, a period of slow-roll inflation
commences, where the Hubble rate is approximately constant.  Also
 plotted are the curvature scale (blue dashed curve), the current Hubble
 scale (red dotted curve) and the scale
 associated with the onset of suppression (green long-dashed curve). The
 points at which these curves intersect the solid black curve corresponds to the
 horizon-crossing times. The shaded region corresponds to the fast-roll period.   
}\label{three_stages}
\end{figure}

As confirmed in the toy models of the preceding section, in general we expect
that the dynamics can be considered in three stages.  The first is the
curvature-dominated stage, where $H = 1/a$, such that $a = t$ and
$\dot{\phi}\simeq -V_\phi t/4$.  During this stage the Hubble friction
term is large, meaning that $\dot\phi$ is very small.  After the time
$t_\ast$, defined by $V(\phi_\ast)/3=1/a^2_\ast$, the potential energy
comes to dominate over the curvature term.  In the initial
potential-dominated stage the slope of the potential may still be
somewhat steeper than usually dictated by the requirement for slow roll.
As such, in this stage we might expect $\dot\phi$ to be somewhat
enhanced, leading to a suppression of the scalar power spectrum for
modes leaving the horizon during this period.  Finally, as the field
evolves away from the barrier, the potential flattens and usual
slow-roll inflation occurs.  The above scenario is depicted
schematically in figure
\ref{three_stages}.

In addition to the evolution of the Hubble parameter, figure
\ref{three_stages} shows the evolution of the physical length
scales associated with the curvature scale, $p_{\rm K} = 1$, the current
Hubble scale, $p_{H_0}$, and
the scale associated with the onset of power suppression, $p_{\rm sup}$.  These scales can all be constrained using observations, which
allows us to determine the relative position of the corresponding
curves on the plot.  For a given open
inflation model, we would then require that the time at which the scale
associated with the onset of suppression leaves the horizon coincides
with the onset of slow-roll inflation.

To determine the separation between the curvature scale and the current
Hubble scale, we note that the current observational constraint on
$\Omega_{{\rm K}} \equiv 1/(a_0H_0)^2$ is $\Omega_{\rm K}\lesssim 0.01$.  We therefore
find $\ln(p_{H_0}/p_{{\rm K}}) \gtrsim \ln(10)\simeq 2.3$,
which corresponds to the vertical spacing between the blue and red lines
in figure \ref{three_stages}.

Next, in order to determine an estimate for the separation between the
current Hubble scale and the scale at which the onset of scalar
suppression is observed, we use the results of a recent analysis by
Easther {\it et al.} \cite{Abazajian:2014tqa}.  In this work they
considered a ``broken'' spectrum that is red-tilted up to some given scale
$p_{{\rm sup}}$ and then blue-tilted for larger scales, thus
explaining the suppression on large scales.  The transition is assumed
to take place instantaneously at $p = p_{{\rm sup}}$, and
the best fit is found to be
 $p_{{\rm sup}}/a_0=4.6\times10^{-3}\,{\rm Mpc}^{-1}$, which can be compared
to the value of $p_{H_0}/a_0=(h/3)\times10^{-3}\,{\rm
Mpc^{-1}}\simeq(0.23)\times 10^{-3}\,{\rm Mpc}^{-1}$.  We thus find
$\ln(p_{{\rm sup}}/p_{H_0})\simeq 2.98$, which corresponds
to the vertical separation between the red and green lines in figure \ref{three_stages}.

Combining the above results, we find $\ln(p_{{\rm sup}}/
p_{{\rm K}}) \gtrsim 5.28$, which corresponds to the total vertical separation
between the blue and green lines in figure \ref{three_stages}.  Of
course, we stress that this is a very simple estimate, as defining
and constraining the onset of suppression is non-trivial and
model dependent.  

By considering the evolution of the Hubble parameter during the
transition from curvature domination to slow-roll inflation, we are able
to convert this constraint on the separation in scales to constraints on
the potential.  Let us use the subscripts ``$\ast$'' and ``S'' to denote
quantities at the time of curvature-potential equality and the onset of slow-roll   
inflation respectively.  Then, using the definition of $\epsilon \equiv
d\ln H/dN$, we have 
\begin{equation}
 H_S = H_\ast \exp\left[\int^S_\ast \epsilon dN\right].
\end{equation}
By definition, we also have $a_S = a_\ast \exp\left[\Delta N\right]$,
where $\Delta N$ is the number of e-foldings between curvature-potential
equality and the onset of slow-roll inflation. As such, for scales associated
with the onset of power suppression, i.e. scales leaving the horizon at the
time of the transition between the fast- and slow-roll phases, we
have
\begin{align}
 p_{{\rm sup}} = a_S H_S &= a_\ast H_\ast \exp\left[\Delta N +
						    \int_\ast^S\epsilon
 dN\right]\label{anapp},
\end{align}
where we have made the simplifying assumption that the
horizon-crossing condition $p = aH$ is valid, which should be true
for these scales.  We then note that prior to the time $t_\ast$ we have
curvature domination, during which $aH = 1 = p_{{\rm K}}$.  As
such, we can re-write the above expression as
\begin{align}\nonumber
 \Delta N + \int^S_\ast \epsilon dN &= \ln\left(\frac{p_{{\rm sup}}}{
					  p_{{\rm K}} }\right) \\
&=
 \ln\left(\frac{p_{{\rm sup}}}{p_{H_0}}\right)
 +\ln\left(\frac{1}{\sqrt{\Omega_{{\rm K}}}}\right).
\end{align}
Inserting the numbers discussed above, this gives 
\begin{equation}
 \Delta N + \int^S_\ast \epsilon dN \gtrsim 5.3
\end{equation}

In the case that $\epsilon$ is constant during the fast-roll phase (as
is the case in figure \ref{three_stages}), this
simplifies further to 
\begin{equation}
 \Delta N(1-\epsilon) = \ln\left(\frac{H_\ast}{H_S}\right)\frac{1-\epsilon}{\epsilon} \gtrsim 5.3.
\end{equation}

Another way in which we can try to constrain the form of the potential after
the tunnelling is by looking at how much the large-scale power spectrum
is suppressed.  Taking the naive relation $\mathcal P_{\mathcal R}
\propto H^2/\epsilon$, and further neglecting the variation in $H$, the
ratio of the power spectrum on large scales to that on small scales will
tell us about the ratio of the slow-roll parameters, i.e.
\begin{equation}
 \frac{\mathcal P_{\mathcal R}(p_1)}{\mathcal P_{\mathcal
  R}(p_2)} \simeq
  \frac{\epsilon_{p_2}}{\epsilon_{p_1}} \le 1,
\end{equation}
where $p_1\le p_2$ and $\epsilon_{p_i}$ indicates the value of $\epsilon$ when the scale $p_i$ left the horizon.
In the schematic example depicted in figure \ref{three_stages}, this
approach is particularly simple as $\epsilon$ simply takes on two
different but constant values in the fast- and slow-roll stages.  As we
have seen in Models 1 and 2, however, in more realistic models the
slow-roll parameter is likely to vary with time during the fast-roll
phase, so that the suppression becomes scale dependent.  Furthermore, we
have seen that the fast rolling of the field is not the only source of
suppression for small $p$.  The additional suppression from quantum
tunnelling effects is not captured in this simplified analysis.
  
\section{\label{conclusions}Conclusions}

The latest CMB measurement from WMAP and $Planck$ hint at there being a
deficit in the temperature power spectrum on large scales, and the
possible detection of primordial gravitational waves by the BICEP2 team
only stands to compound this tension.  In light of this, the fact that open inflation
models give rise to a suppression of scalar power on large
scales has led to them receiving renewed interest.
Intuitively, the main source of suppression in these models is
the steepening of the potential towards the barrier that separates the
true and false vacuums.  This steepening leads to a fast-roll phase in which
$\dot\phi$ is enhanced, so that $\mathcal P_{\mathcal R} \propto
H^4/\dot\phi^2$ becomes suppressed.  The requirement that this
fast-rolling should affect observable scales puts a limit on the
duration of ensuing slow-roll inflation.  If inflation does not
last long enough, however, then one risks violating observable
constraints on $\Omega_{{\rm K}}$.  

In this paper we have revisited two of the toy models of single-field open
inflation introduced in \cite{Linde:1999wv}, with potentials given by \eqref{m1pot} and \eqref{pot2}.  We
have seen that both models are indeed capable of giving suppression of
the scalar power on the order of ten percent, and that they are also
able to evade current constraints on $\Omega_{{\rm K}}$.  The steepening
of the potential towards the barrier in Model 1 was of power-law form, and this gradual
steepening meant that the onset of suppression was also gradual,
potentially affecting three orders of magnitude of observable scales.  In Model 2, however,
the steepening was of exponential form, leading to a much shorter
fast-roll phase.  Correspondingly, the range of affected observable
scales was much narrower.     

Finally, in both models we found that in addition to the source of
suppression mentioned above, one also has additional suppression that results
from the system's memory of the tunnelling phase. (See also \cite{Linde:1999wv}.)  In the
case that current bounds on $\Omega_{{\rm K}}$ are almost saturated,
we saw that this additional suppression can be non-negligible on
observable scales.  As such, any quantitative analysis would need to
take into account this additional suppression.       

\begin{acknowledgments}
We would like to thank R.\,Saito, K.\,Sugimura, T.\,Tanaka and D.\,Yeom for valuable
discussions. This work was supported by the JSPS Grant-in-Aid for
Scientific Research (A) No.~21244033.
\end{acknowledgments}

\bibliography{openbib}{}

\end{document}